\documentclass[12pt]{article}
\usepackage{graphicx}
\def \bea{\begin{eqnarray}}
\def \beq{\begin{equation}}

\def \eea{\end{eqnarray}}
\def \eeq{\end{equation}}

\renewcommand{\thetable}{\Roman{table}}
\begin{document}
\begin{flushright}
EFI 06-20 \\
hep-ph/0609195 \\
September 2006 \\
\end{flushright}

\renewcommand{\thetable}{\Roman{table}}
\centerline{\bf HADRON SPECTROSCOPY:  THEORY AND EXPERIMENT}
\medskip

\centerline{Jonathan L. Rosner}
\centerline{\it Enrico Fermi Institute and Department of Physics,
  University of Chicago}
\centerline{\it 5640 S. Ellis Avenue, Chicago IL 60637 USA}

\begin{quote}
Many new results on hadron spectra have been appearing in the past few years
thanks to improved experimental techniques and searches in new channels.  New
theoretical techniques including refined methods of lattice QCD have kept pace
with these developments.  Much has been learned about states made of both light
($u,~d$, and $s$) and heavy ($c,~b$) quarks.  The present review treats
light-quark mesons, glueballs, hybrids, particles with a single $c$ or $b$
quark, charmonium ($c \bar c$), and bottomonium ($b \bar b$) states.  Some
prospects for further study are noted.
\end{quote}

\leftline{PACS numbers: 14.20.Lq, 14.40.Cs, 14.40.Gx, 14.40.Lb}

\section{INTRODUCTION \label{sec:intro}}

Quantum Chromodynamics (QCD) is our theory of the strong interactions:
Its property of asymptotic freedom \cite{AF} allows one to interpret deep
inelastic scattering experiments \cite{DIS} in terms of pointlike nucleon
constituents \cite{part}, and permits perturbative calculations of
short-distance phenomena.  However, we are far from understanding how it works
at longer distances governing the spectra of hadrons.  Many hadrons discovered
recently have unexpected properties.  We need to understand hadron spectra
in order to separate electroweak physics from strong-interaction effects.
Moreover, we may need to use our experience with QCD in dealing with any
non-perturbative effects encountered at higher energies such as those to be
probed by the Large Hadron Collider (LHC).  The understanding of electroweak
symmetry breaking may well require non-perturbative techniques at TeV scales
similar to those useful for hadron spectroscopy at GeV scales.  Sharpening
spectroscopic techniques even may help understand the intricate structure of
masses and transitions, itself the indication of a rich spectroscopy at the
quark and lepton level.

The QCD scale is $\sim 200$ MeV (momentum) or $\sim 1$ fm (distance), where
perturbation theory cannot be used.  Although lattice gauge theories are, in
principle, the way to describe effects in this regime, several other methods
can provide information, especially for multi-quark and multi-hadron problems
not yet feasible with lattice techniques.  Some of these will be discussed
briefly in Section \ref{sec:meth}.  I will then describe some phenomena to
which these methods
can be applied.  These include light-quark and no-quark mesons (Section
\ref{sec:lq}), charmed (Section \ref{sec:charm}) and beauty/bottom (Section
\ref{sec:beauty}) hadrons, and heavy charmonium (Section \ref{sec:charmon}) and
bottomonium (Section \ref{sec:upsilons}).  I conclude with some prospects for
progress in the areas mentioned (Section \ref{sec:future}) and summarize in
Section \ref{sec:summary}.  Baryon spectroscopy is the topic of a separate
review in this issue \cite{LB}.  An extended review of transitions in
quarkonium, drawing on and extending some of the material in this article, may
be found in Ref.\ \cite{Eichten:2007qx}.

\section{SOME THEORETICAL METHODS \label{sec:meth}}

\subsection{Lattice gauge theories}
\vskip -0.1in

At momentum scales less than about 2 GeV/$c$ the QCD coupling constant
$\alpha_S(Q^2)$ becomes large enough that perturbation theory cannot be used.
[For recent compilations of evidence that $\alpha_S(Q^2)$ behaves as
predicted by asymptotic freedom see Ref.\ \cite{Bethke:2006ac}.]  Below this
scale one must resort to non-perturbative methods to describe long-distance
hadronic interactions.

If space-time is discretized, one can overcome the dependence in QCD on
perturbation theory.  Quark confinement is one consequence of this lattice
gauge theory approach.  An accurate description of the heavy quarkonium
spectrum can be obtained once one takes account of degrees of freedom
associated with the production of pairs of light ($u,d,s$) quarks
\cite{Davies:2003ik}.

\subsection{Chiral dynamics, unitarity, and crossing symmetry}
\vskip -0.1in

The coupling of soft pions to hadronic systems is specified by PCAC (partial
conservation of axial current) and current algebra \cite{Weinberg:1966kf}.
When supplemented by constraints enforcing unitarity and crossing symmetry,
current algebra becomes a powerful tool for describing the dynamics of
mesons up to the GeV scale and baryons somewhat higher.  An early application
of the power of these principles to $\pi \pi$ scattering appears in Ref.\
\cite{BG71}.  Most treatments these days implement crossing symmetry using an
elegant set of exact low-energy relations \cite{Roy71}.  A systematic
expansion in chiral loops is known as {\it chiral perturbation theory} and
is described in detail in Refs.\ \cite{Gasser:1983yg}.  (For a modern review,
see Ref.\ \cite{Bernard}.)

Although chiral $L \times R$ symmetry is usually regarded as being
spontaneously broken to the vector symmetry, with the symmetry breaking
manifested through the existence of Nambu-Goldstone bosons like the pion, 
a Wigner-Weyl realization of the chiral symmetry would be visible through
parity doubling of states.  Some evidence for this in baryon spectra has
been noted \cite{Jaffe:2005aq}.

\subsection{Heavy quark symmetry}
\vskip -0.1in

Hadrons with one charmed or beauty quark can be regarded as ``atoms'' of QCD,
with the light-quark and gluonic degrees of freedom playing the role of the
electron(s) and the heavy quark playing the role of the nucleus.  Properties
of these systems tend to be very simple under the interchange $c
\leftrightarrow b$, in the manner of isotope effects in nuclei.  These
features were first noted in Ref.\ \cite{Nussinov:1986hw} and systematically
explored in Refs.\ \cite{Isgur:1989vq}.  One can combine heavy quark symmetry
with chiral dynamics to predict the existence of a broken form of parity
doubling in mesons containing a single heavy quark \cite{Nowak:1992um}.  We
will see that there is some evidence for this in charmed-strange mesons.

\subsection{Quark model potential descriptions}
\vskip -0.1in

Quarkonium states may be described as bound by a force whose
short-distance behavior is approximately Coulombic (with a slow variation of
coupling strength to account for asymptotic freedom) and whose long-distance
behavior is linear to account for quark confinement \cite{Cornell}.  (For
early reviews see Refs.\ \cite{Novikov:1977dq,Appelquist:1978aq,Quigg:1979vr,
Grosse:1979xm}).  One can estimate from such a potential (or a simpler
interpolating version \cite{Quigg:1977dd}) that a non-relativistic description
for charmonium is quite crude, with characteristic velocities $\langle v^2
\rangle \simeq 0.5$ for a charmed quark in a $c \bar c$ bound state.  For a $b$
quark in a $b \bar b$ bound state $\langle v^2 \rangle \simeq 0.15$, so the
non-relativistic description begins to make some sense.

The partial widths for a $^3S_1$ state to decay to a lepton pair through a
virtual photon depend on the square of the relative wave function at the
origin through the relation \cite{VanRoyen:1967nq}
\beq
\Gamma(^3S_1 \to e^+ e^-) = \frac{16 \pi \alpha^2 e_Q^2 |\Psi(0)|^2}{M^2}~~~,
\eeq
where $e_Q = 2/3$ or $-1/3$ is the quark charge and $M$ is the mass of the
$^3S_1$ state.  Thus leptonic partial widths probe the compactness of
the quarkonium system, and provide important information complementary to
level spacings.  In a power-law potential $V(r) \sim {\rm sgn}(\nu) r^\nu$,
$|\Psi(0)|^2$ scales as $m_Q^{3/(2+\nu)}$, or $\sim m_Q^3, m_Q^{3/2}, m_Q$ for
$\nu=-1,0,1$ \cite{Quigg:1979vr}.  Thus the effective quark mass in a potential
description is constrained by measured leptonic widths.  In more fundamental
descriptions such as lattice gauge theories similar constraints will hold.

Hyperfine and fine-structure splittings in quarkonium are sensitive to
the Lorentz structure of the interquark interaction 
\cite{Novikov:1977dq,Appelquist:1978aq,KQR,Brambilla:2004wf}.
Writing the effective potential $V(r)$ as the sum of Lorentz vector $V_V$ and
Lorentz scalar $V_S$ contributions, one finds that the spin-spin interaction
is due entirely to the Lorentz vector:
\beq
V_{SS}(r) = \frac{\sigma_1 \cdot \sigma_2}{6 m_Q^2} \nabla^2 V_V(r)~~~,
\eeq
where $\sigma_1$ and $\sigma_2$ are Pauli matrices acting on the spins of
the quark and antiquark, respectively.  For a Coulomb-like potential $\sim
-1/r$ the Laplacian is proportional to $\delta^3(r)$, so that $V_{SS}(r)$
contributes to hyperfine splittings only for S waves, whose wave functions
are non-zero at the origin.  In QCD the coupling constant undergoes slow
(logarithmic) variation with distance, leading to small non-zero contributions
to hyperfine splittings for $L > 0$ states.

Spin-orbit and tensor forces affect states with $L > 0$.
The spin-orbit potential is
\beq
V_{LS}(r) = \frac{L \cdot S}{2 m_Q^2 r}
 \left(3 \frac{dV_V}{dr} - \frac{dV_S}{dr} \right)~~~,
\eeq
where $L$ is the relative orbital angular momentum of $Q$ and $\bar Q$, while
$S$ is the total quark spin.  The tensor potential is
\beq
V_T(r) = \frac{S_{12}}{12 m_Q^2} \left( \frac{1}{r} \frac{dV_V}{dr}
 - \frac{d^2V_V}{dr^2} \right)~~~,
\eeq
where $S_{12} \equiv 2[3(S \cdot \hat{r})(S \cdot \hat{r}) - S^2]$ has
non-zero expectation values only for $L >0$.

One must include relativistic effects to address the high quality of quarkonium
data now available.  Early treatments include Refs.\ \cite{Godfrey:1985xj} and
\cite{Capstick:1986bm} (with a recent application in \cite{Godfrey:2004ya}).
The papers of Refs.\ \cite{Ebert:2002pp} represent another area of effort in
this regard.  However, a recent review of relativistic corrections to electric
dipole matrix elements \cite{Eichten:2007qx} indicates that a satisfactory
description of these properties of hadrons has not yet been reached.

Coupled-channel effects are also very important in heavy quarkonium
spectroscopy, particularly for states near and above flavor threshold.
The study of these was pioneered in Ref.\ \cite{Eichten:1978tg} (for
recent applications see \cite{Eichten:2004uh}).

\subsection{Light quarks as quasi-particles}
\vskip -0.1in

One can reproduce a great deal of the spectrum of hadrons containing the
$u,~d$, and $s$ quarks with a simple model based on additive quark masses
$m_i$ and hyperfine interactions proportional to $\langle \sigma_i \cdot
\sigma_j /(m_i m_j) \rangle$ \cite{DeRujula:1975ge,Sakharov:1980ph,%
Gasiorowicz:1981jz}.  In this treatment the best fit to meson spectra occurs
when $u$ and $d$ quarks have effective masses of 310 MeV/$c^2$ while $s$
quarks have effective masses of 485 MeV/$c^2$.  The corresponding values for
baryons are each shifted upward by 53 MeV/$c^2$:  363 and 538 MeV/$c^2$.  These
values are very different from the effective masses of quarks at $Q^2$ scales
of (2 GeV/$c^2)^2$, which are about 3, 6, and 100 MeV/$c^2$ \cite{PDG}.  The
additional mass in the ``constituent'' quarks may be thought of as due to
their interaction with the surrounding gluon field.

\subsection{Correlations among quarks}
\vskip -0.1in

Because the product of two color-SU(3) triplets consists of an antitriplet
and a sextet, two color-triplet quarks can form a color antitriplet with
spin zero or 1.  The spin-spin force is attractive in the spin-zero state,
so two non-identical quarks can form a spinless boson whose effective mass
can be comparable to that of a single quark.  An early realization that this
fact implied a sort of supersymmetry in the hadron spectrum appears in
Refs.\ \cite{Roncaglia:1994ex}.  Recent applications of this idea to hadron
spectroscopy have appeared in Refs.\ \cite{KL,JW,SW}.  Quantitative tests of
these ideas include the prediction \cite{KL} of a weakly decaying $bq \bar c
\bar q'$ state.

\subsection{QCD sum rules and instantons}
\vskip -0.1in

Because of its increased coupling strength at long distances, QCD leads to
the formation of {\it condensates} (similar to Cooper pairs in
superconductivity), including non-zero expectation values of color singlet
quark-antiquark pairs and gluonic configurations such as {\it instantons}.
A systematic attempt to cope with the effect of these condensates on hadron
spectroscopy relies on sum rules pioneered in Ref.\ \cite{Shifman:1978bw}.  For
a more recent review of the role of instantons in hadron spectroscopy, see
Ref.\ \cite{Schafer:1996wv}.

\subsection{Resonance decays}
\vskip -0.1in

An elegant treatment of hadronic resonance decays in which a
single quark in a hadron undergoes pion emission was given in Ref.\
\cite{Gilman:1973hc}.  This work and similar treatments of electromagnetic
decays \cite{Gilman:1973jm,Hey:1974qe} are based on the work of Melosh
\cite{Melosh:1974cu} relating ``current quarks'' (those in the fundamental
QCD Lagrangian) to ``constituent quarks'' (essentially quasi-particles in
terms of which hadrons have simple properties).

From the standpoint of predictions the above single-quark-transitions is
identical to a model in which resonance decays proceed via the creation
of a $^3P_0$ quark-antiquark pair with the quantum numbers of the vacuum,
$J^{PC} = 0^{++}$ \cite{Micu:1968mk,Colglazier:1970vx,Petersen:1972qk,%
LeYaouanc:1972ae}.  In resonance decays involving two possible partial waves,
these approaches involve independent amplitudes for each partial wave.
Models based on explicit quark wave functions (e.g., \cite{Feynman:1971wr,%
Isgur:1978xj}) relate the partial waves to one another.

\section{LIGHT-QUARK STATES\label{sec:lq}}

In this section we discuss a few topics of current interest in light-quark
spectroscopy: the nature of the low-energy S-wave $\pi \pi$ and $K \pi$
interactions; the proliferation of interesting threshold effects in a
variety of reactions, and the interaction of quark and gluonic degrees
of freedom.
\vskip -0.3in

\subsection{Low-energy $\pi \pi$ S-wave}
\vskip -0.1in

An S-wave $\pi \pi$ low-mass correlation in the $I=0$ channel (``$\sigma$'')
has been used for many years to describe nuclear forces.  Is it a resonance?
What is its quark content?  What can we learn about it from charm and beauty
decays?  This particle, otherwise known as $f_0(600)$ \cite{PDG}, can be
described as a dynamical $I=J=0$ resonance in elastic $\pi \pi$ scattering
using current algebra, crossing symmetry, and unitarity \cite{BG71,VB85,%
Dobado:1992ha}.  It appears as a pole with a large imaginary part with real
part at or below $m_\rho$.  Its effects differ in $\pi \pi \to \pi \pi$, where
an Adler zero suppresses the low-energy amplitude, and inelastic processes
such as $\gamma \gamma \to \pi \pi$ \cite{GR}, where the lack of an Adler
zero leads to larger contributions at low $m_{\pi \pi}$.

There is not unanimity regarding the exact position of the $\sigma$ pole.  In
one approach \cite{CCL} it is found at $441 - i 272$ MeV, corresponding to a
full width at half maximum of 544 MeV; another \cite{vanBeveren:2006ua} finds
it at $555 - i 262$ MeV.  Such a $\sigma$ provides a good description of
$\gamma \gamma \to \pi^0 \pi^0$ \cite{Pennington:2006dg}, with $\Gamma(\sigma
\to \gamma \gamma) = (4.1 \pm 0.3)$ keV.  While this large partial width might
be viewed as favoring a $q \bar q$ interpretation of $\sigma$
\cite{Pennington:2006dg}, a $\pi \pi$ dynamical
resonance seems equally satisfactory \cite{GR}.  Other recent manifestations
of a $\sigma$ include the decays $D^+ \to \sigma \pi^+ \to \pi^+ \pi^- \pi^+$
\cite{E791sigma} and $J/\psi \to \omega \sigma \to \omega \pi^+ \pi^-$
\cite{BESsigma}, where the $\sigma$ pole appears at $(541\pm 39) - i
(252 \pm 42)$ MeV (or $(500 \pm 30) - i(264 \pm 30)$ MeV in an independent
analysis \cite{Bugg:2006gc}).  Successful fits without a $\sigma$ have been
performed, but have been criticized in Ref.\ \cite{Bugg:2005nt}.
\vskip -0.3in

\subsection{Low-energy $K \pi$ S-wave}
\vskip -0.1in

Is there a low-energy $K \pi$ correlation (``$\kappa$'')?  Can it be
generated dynamically in the same manner as the $\sigma$?  Some insights are
provided in \cite{Dobado:1992ha,Oller}.

The low-energy $K \pi$ interaction in the $I=1/2,~J=0$ channel is favorable to
dynamical resonance generation:  The sign of the scattering length is the
same as for the $I=J=0~\pi \pi$ interaction.  A broad scalar resonance $\kappa$
is seen in the $I=1/2,~J=0~K^- \pi^+$ subsystem in $D^+ \to K^-\pi^+  \pi^+$,
and a model-independent phase shift analysis shows resonant $J=0$ behavior in
this subsystem \cite{Aitala:2002kr}.  The $\kappa$ is also seen by the BES II
Collaboration in $J/\psi \to \bar K^{*0}(892) K^+ \pi^-$ decays
\cite{Ablikim:2005ni}.  An independent analysis of the BES II data
\cite{vanBeveren:2006ua} finds a $\kappa$ pole at $745 - i 316$ MeV, while a
combined analysis of $D^+ \to K^-\pi^+  \pi^+$, elastic $K \pi$ scattering,
and the BES II data \cite{Bugg:2005xx} finds a pole at $M(\kappa) =
(750^{+30}_{-55}) - i(342 \pm 60)$ MeV.


The $\kappa$, like the $\sigma$, is optional in many descriptions of
final-state interactions.  For example, in a recent fit to the $D^0 \to K^+ K^-
\pi^0$ Dalitz plot based on CLEO data \cite{Paras06}, bands are seen
corresponding to $K^{*-}$, $K^{*+}$, and $\phi$.  One can see the effect of an
S-wave amplitude interfering with $K^{*+}$ and $K^{*-}$ with opposite signs in
the low-$K^\pm \pi^0$-mass regions of the plot, but one cannot tell whether
this amplitude is non-resonant or due to a $\kappa$.
Depopulated regions at $m(K^\pm \pi^0) \simeq$ 1 GeV/$c^2$ may be due to the
opening of the $K \pi^0 \to K \eta$ S-wave threshold (a $D^0 \to K^+ K^- \eta$
Dalitz plot would test this) or to a vanishing S-wave $K \pi$ amplitude
between a $\kappa$ and a higher $J^P = 0^+$ resonance.  A candidate for such
a resonance exists around 1430 MeV/$c^2$ \cite{PDG}.

\subsection{Dips and edges}
\vskip -0.1in

High-statistics Dalitz plots for heavy meson decays often exhibit dips and
edges associated with thresholds \cite{Rosner:2006vc}.  In a recent $D^0 \to
K_S^0 \pi^+ \pi^-$ plot \cite{BaKspipi}, sharp edges in the $\pi^+ \pi^-$
spectrum correspond to $\rho$-$\omega$ interference [around $M(\pi \pi) = 0.8$
GeV/$c^2$] and to $\pi^+ \pi^- \leftrightarrow K \bar K$ [around $M(\pi \pi) =
1$ GeV/$c^2$].  Rapid variation of an amplitude occurs when a new S-wave
channel opens because no centrifugal barrier is present.

Further dips are seen in $6 \pi$ photoproduction just at $p \bar p$ threshold;
in $R_{e^+ e^-}$ just below the threshold for S-wave production of $D(1865)
+D_1(2420)$; and in the Dalitz plot for $B^\pm \to K^\pm K^\mp K^\pm$
around $M(K^+ K^-) = 1.6$ GeV/$c^2$ \cite{Aubert:2006nu}, which could be a
threshold for vector meson pair production.

\subsection{Glueballs and hybrids}
\vskip -0.1in

In QCD, quarkless ``glueballs'' may be constructed from color-singlet
polynomials in the field-strength tensor $F_{\mu \nu}^a$.  All such states
should be flavor-singlet with isospin $I=0$, though $s \bar s$ couplings of
spinless states could be favored \cite{Chanowitz:2005du}.  Lattice QCD
predicts the lowest glueball to be $0^{++}$ with $M \simeq 1.7$ GeV
\cite{Campbell:1997}.  The next-lightest states, $2^{++}$ and $0^{-+}$, are
expected to be several hundred MeV/$c^2$ heavier.  Thus it is reassuring
that the lightest mainly flavor-singlet state, the $\eta'$, is only gluonic
$(14 \pm 4)\%$ of the time, as indicated by a recent measurement of ${\cal B}
(\phi \to \eta' \gamma)$ by the KLOE Collaboration \cite{Ambrosino:2006gk}.

Many other $I=0$ levels, e.g., $q \bar q$, $q \bar q g$ ($g$ = gluon), $q q
\bar q \bar q, \ldots$, can mix with glueballs.  One must study $I=0$ levels
and their mesonic couplings to separate out glueball, $n \bar n \equiv (u \bar
u + d \bar d)/ \sqrt{2}$, and $s \bar s$ components.  Understanding the rest of
the {\it flavored} $q \bar q$ spectrum for the same $J^P$ thus is crucial.
The best $0^{++}$ glueball candidates (mixing with $n \bar n$ and $s \bar s$)
are at 1370, 1500, and 1700 MeV.  One can explore their flavor structure
through production and decay, including looking for their $\gamma(\rho,\omega,
\phi)$ decays \cite{ClZh}.  A CLEO search for such states in $\Upsilon(1S)
\to \gamma X$ finds no evidence for them but does see the familiar resonance
$f_2(1270)$ \cite{CLf}.

QCD predicts that in addition to $q \bar q$ states there should be $q \bar q g$
(``hybrid'') states containing a constituent gluon $g$.  One signature of them
would be states with quantum numbers forbidden for $q \bar q$ but allowed for
$q \bar q g$.  For $q \bar q$, $P = (-1)^{L+1},~C = (-1)^{L+S}$, so $CP =
(-1)^{S+1}$.  The forbidden $q \bar q$ states are then those with $J^{PC}=
0^{--}$ and $0^{+-},~1^{-+},~2^{+-},\ldots$.  In quenched lattice QCD most
calculations find that the lightest exotic hybrids have $J^{PC} = 1^{-+}$ and
$M(n \bar n g) \simeq 1.9$ GeV, $M(s \bar s g) \simeq 2.1$ GeV, with errors
0.1--0.2 GeV \cite{McNeile:2002az}.  (Ref.\ \cite{Hedditch:2005zf} finds that
such estimates, based on linear extrapolations
to the chiral limit, may have overestimated these masses by as much as
0.3 GeV/$c^2$.  Unquenched QCD must treat mixing with $qq \bar q \bar
q$ and meson pairs.) Candidates for hybrids include $\pi_1(1400)$ (seen in some
$\eta \pi$ final states, e.g., in $p \bar p$ annihilations) and $\pi_1(1600)$
(seen in $3 \pi$, $\rho \pi$, $\eta' \pi$).  While Brookhaven experiment E-852
published evidence for a $1^{-+}$ state called $\pi_1(1600)$
\cite{Adams:1998ff}, a recent analysis
\cite{Dzierba:2005sr} does not require this particle if a $\pi_2(1670)$
contribution is assumed instead.  The favored decays of a $1^{-+}$ hybrid are
to a $q \bar q (L=0)$ + $q \bar q (L=1)$ pair, such as $\pi b_1(1235)$.  A
review of glueballs and hybrids has been presented recently \cite{Meyer}, and
lattice predictions for their decays have been discussed in Ref.\
\cite{Michael:2006hf}.

\section{CHARMED STATES\label{sec:charm}}

\begin{figure}
\includegraphics[height=0.44\textheight]{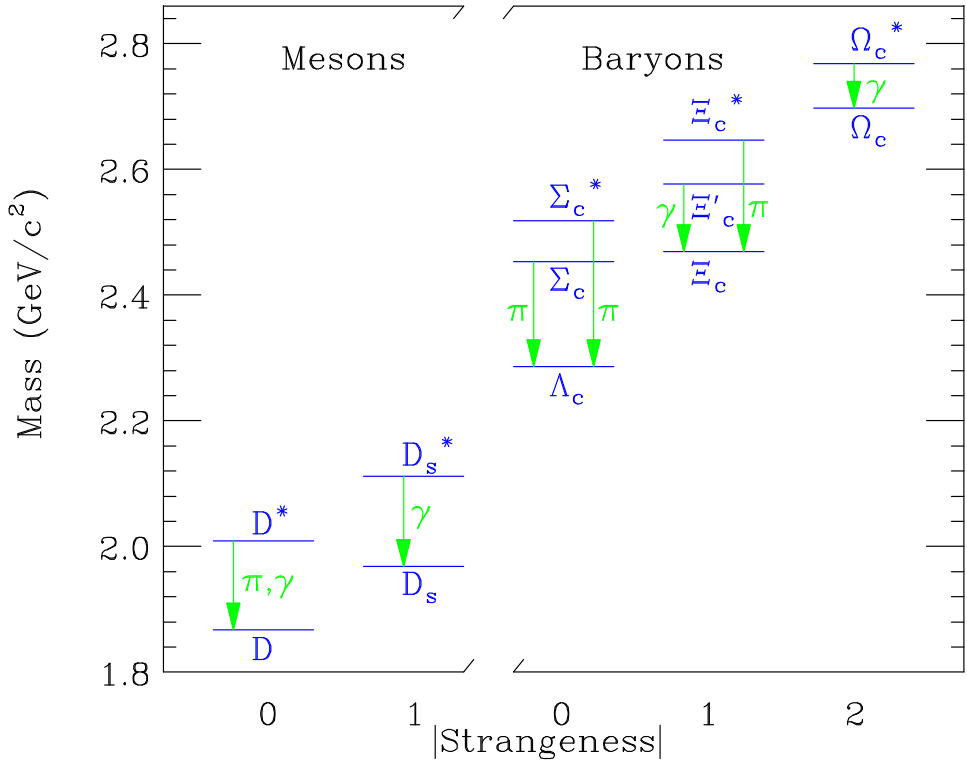}
\caption{Lowest S-wave states with a single charmed quark.
\label{fig:charm}}
\end{figure}

The present status of the lowest S-wave states with a single charmed quark is
shown in Fig.\ \ref{fig:charm}.  We will discuss progress on charmed baryons,
on charmed-strange mesons, and on $D^+$ and $D_s^+$ decay constants.
\vskip -0.3in

\subsection{Charmed baryons}
\vskip -0.1in

The predicted spectroscopy of ground state baryons containing a single charmed
quark was mapped out in 1974 \cite{Gaillard:1974mw}.  At last, the family of
predicted states has been completed with the report by the BaBar Collaboration
of the $\Omega_c^*$ \cite{Aubert:2006je}.  It is a candidate for the $J^P =
3/2^+$ state of the $css$ system.  It was found to lie $70.8 \pm 1.0 \pm 1.1$
MeV/$c^2$ above the $\Omega_c$ (the candidate for the S-wave $J^P = 1/2^+$ $css$
state), and to decay to $\gamma \Omega_c$.  Its mass was as expected in
models of QCD hyperfine splittings (e.g., \cite{Rosner:1995yu}).

\begin{figure}
\includegraphics[height=0.43\textheight]{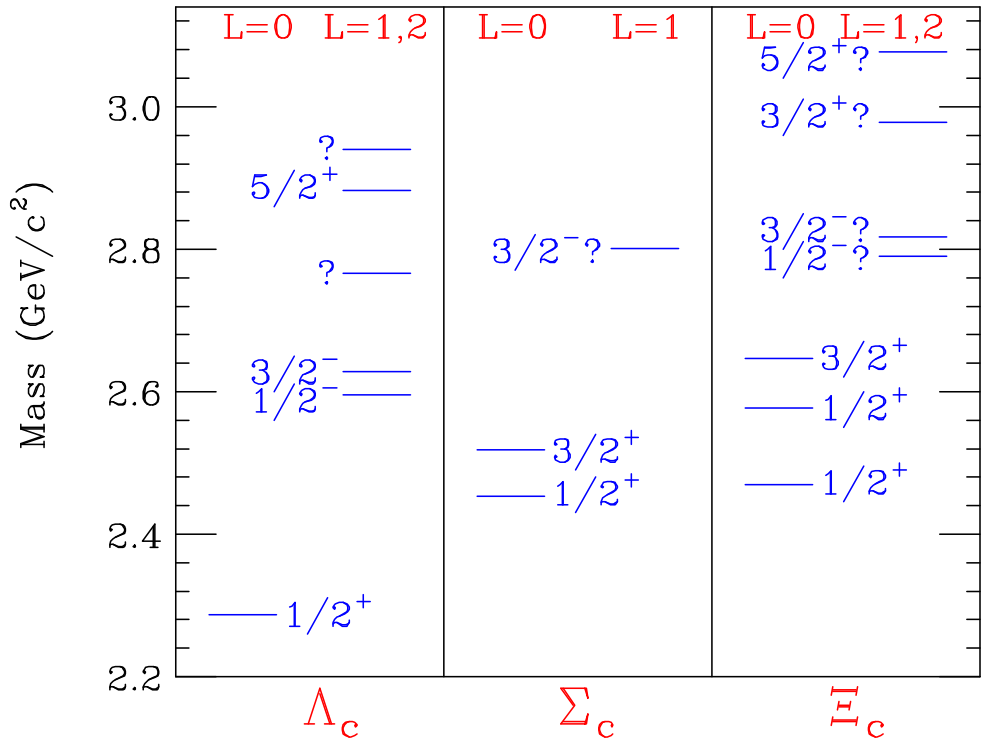}
\caption{Singly-charmed baryons and some of their orbital excitations.
\label{fig:lexb}}
\end{figure}

CLEO, BaBar, and Belle have contributed greatly to data on orbitally-excited
charmed baryons, whose levels are plotted along with those of the lowest $L=0$
states in Fig.\ \ref{fig:lexb}.  The first excitations of the $\Lambda_c$ and
$\Xi_c$ are similar, scaling well from the first $\Lambda$ excitations
$\Lambda(1405,1/2^-)$ and $\Lambda(1520,3/2^-)$.  They have the same cost in
$\Delta L$ (about 300 MeV), and their $L \cdot S$ splittings scale as $1/m_s$
or $1/m_c$.  Higher $\Lambda_c$ states may correspond to excitation of a
spin-zero $[ud]$ pair to $S=L=1$, leading to many allowed $J^P$ values up to
$5/2^-$.  In $\Sigma_c$ the light-quark pair has $S=1$; adding $L=1$ allows
$J^P \le 5/2^-$.  States with higher $L$ may be narrower as a result of
increased barrier factors affecting their decays, but genuine spin-parity
analyses would be very valuable.  Here are some recent results:

(1) The $\Lambda_c(2880)$ was first seen in the $\Lambda_c^+ \pi^- \pi^+$ mode
by CLEO \cite{Artuso:2000xy} and confirmed in the $D^0 p$ mode recently by
BaBar \cite{Aubert:2006sp}.  It now has been shown to have likely $J^P = 5/2^+$
\cite{Abe:2006rz}, on the basis of the angular distribution in its
$\Sigma_c(2455) \pi$ decays and the small measured ratio
$\Gamma[\Sigma_c(2520) \pi]/\Gamma[\Sigma_c(2455) \pi] \simeq 1/4$.

(2) The highest $\Lambda_c$ was seen by BaBar in the decay mode $D^0 p$
\cite{Aubert:2006sp}.  The Belle Collaboration has seen evidence for its
decay to $\Sigma_c(2455) \pi$ \cite{Abe:2006rz}.

(3) An excited $\Sigma_c$ candidate has been seen decaying to $\Lambda_c \pi^+$,
with mass about 510 MeV above $M(\Lambda_c)$ \cite{Mizuk:2004yu}.  Its $J^P$
shown in Fig.\ \ref{fig:lexb} is a guess, using ideas of \cite{SW}.

(4) The highest $\Xi_c$ levels were reported by the Belle Collaboration in
Ref.\ \cite{Chistov:2006zj}, and confirmed by BaBar \cite{Aubert:2006uw}.
Both are seen in the $\Lambda_c^+ K^- \pi^+$ channel.  In the Belle data, a
state with $M = 2978.5 \pm 2.1 \pm 2.0$ MeV/$c^2$ has width $\Gamma = 43.5 \pm
7.5 \pm 7.0$ MeV, while one with $M = 3076.7 \pm 0.9 \pm 0.5$ MeV/$c^2$ has
width $\Gamma = 6.2 \pm 1.2 \pm 0.8$ MeV.  The isospin partner of the higher
state is also seen at $M = 3082.8 \pm 1.8 \pm 1.5$ MeV/$c^2$ in the
$\Lambda_c^+ K_S \pi^-$ mode.  The BaBar data are qualitatively similar, with
the lower state having a width of $(23.6 \pm 2.8 \pm 1.3)$ MeV and the higher
having a width of $(6.2 \pm 1.6 \pm 0.5)$ MeV.  The masses of these states
suggest that they could have $L=2$, hence positive parity.  If the light quarks
in these states are coupled to spin 0, as expected for the lowest orbital
excitations, the allowed $J^P$ values are $3/2^+$ and $5/2^+$.  The disparity
of the total widths suggests that the higher of the two may have $J^P = 5/2^+$.
Its decay to $\Sigma_c^{++} K^-$ would then have to be via an F-wave.  (The
$\Sigma_c^{*++} K^-$ channel, seen by BaBar \cite{Aubert:2006uw}, would be
available to a P-wave.) If the lower state had $J^P = 3/2^+$, it could decay to
$\Sigma_c^{++} K^-$ via a P-wave, and hence be broader.

\subsection{Excited charmed-strange states}
\vskip -0.1in

\begin{figure}
\includegraphics[width=0.97\textwidth]{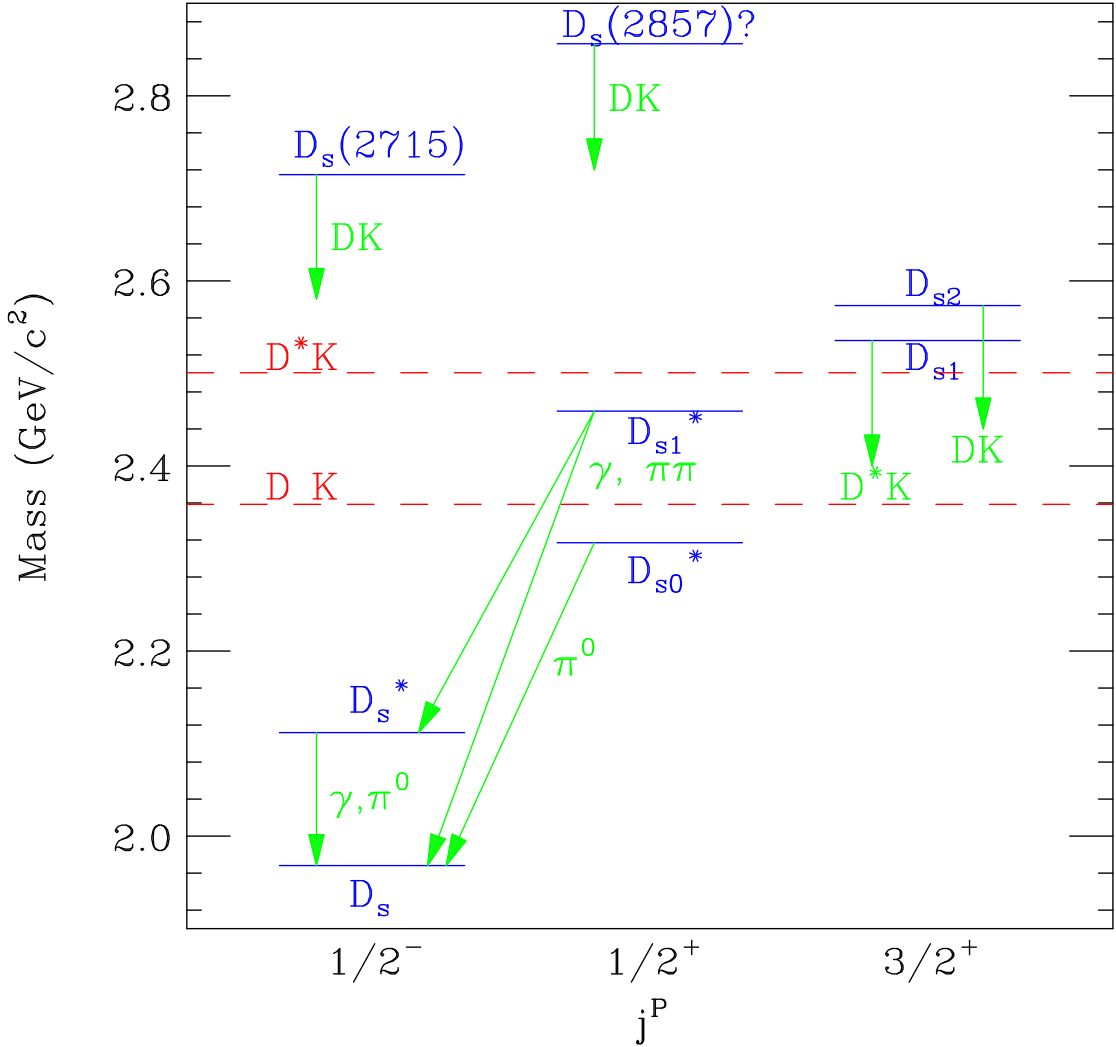}
\caption{Charmed-strange mesons with $L=0$ (negative-parity), $L=1$
(positive-parity), and candidate for state with $L=2$ (positive parity).  Here
$j^P$ denotes the total light-quark
spin + orbital angular momentum and the parity $P$.
\label{fig:ds}}
\end{figure}

In the past couple of years the lowest $J^P = 0^+$ and $1^+$ $c \bar s$
states turned out to have masses well below most expectations.  If they had
been as heavy as the already-seen $c \bar s$ states with $L=1$, the
$D_{s1}(2536)$ [$J^P = 1^+$] and $D_{s2}(2573)$ [$J^P = 2^+$]), they would
have been able to decay to $D \bar K$ (the $0^+$ state) and $D^* \bar K$ (the
$1^+$ state).  Instead several groups \cite{Aubert:2003fg} observed a narrow
$D_s(2317) \equiv D_{s0}^*$ decaying to $\pi^0 D_s$ and a narrow $D_s(2460)
\equiv D_{s1}^*$ decaying to $\pi^0 D_s^*$, as illustrated in Fig.\
\ref{fig:ds}.  Their low masses allow the isospin-violating and electromagnetic
decays of $D_{s0}^*$ and $D_{s1}^*$ to be observable.  The decays $D_s(2460)
\to D_s \gamma$ and $D_s(2460) \to D_s \pi^+ \pi^-$ also have been seen
\cite{Marsiske,Aubert:2006nm}, and the absolute branching ratios
${\cal B}(D_{s1}^* \to \pi^0 D_s^*) = 0.56 \pm 0.13 \pm 0.09,$
${\cal B}(D_{s1}^* \to \gamma D_s) = 0.16 \pm 0.04 \pm 0.03,$
${\cal B}(D_{s1}^* \to \pi^+ \pi^- D_s^*) = 0.04 \pm 0.01$
measured.

The selection rules in decays of these states show that their $J^P$ values
are consistent with $0^+$ and $1^+$.  Low masses are predicted
\cite{Nowak:1992um} if these states are viewed as parity-doublets of the
$D_s(0^-)$ and $D^*_s(1^-)$ $c \bar s$ ground states in the framework of
chiral symmetry.  The splitting from the ground states is 350 MeV in each case.
Alternatively, one can view these particles as bound states of $D^{(*)}K$,
perhaps bound by the transitions $(c \bar q)(q \bar s) \leftrightarrow (c \bar
s)$ (the binding energy in each case would be 41 MeV), or as $c \bar s$ states
with masses lowered by coupling to $D^{(*)}K$ channels
\cite{vanBeveren:2003kd,Close:2004ip}.

A candidate for the first radial excitation of the $D_s^*(2112)$ has been
observed in $B^+ \to \bar D^0 D^0 K^+$ decays \cite{Abe:2006xm} in the
$M(D^0 K^+)$ spectrum.  Its mass and width are $(2715 \pm 11^{+11}_{-14})$ and
$(115 \pm 20^{+36}_{-32})$ MeV/$c^2$.  Its spin-parity is $J^P = 1^-$.  It lies
$(603^{+16}_{-18})$ MeV/$c^2$ above the ground state, in between the
$2^3S_1$--$1^3S_1$ spacings of $(681 \pm 20)$ MeV/$c^2$ for $s \bar s$
and 589 MeV/$c^2$ for $c \bar c$ \cite{PDG}.  This is as expected in an
interpolating
potential such as a power-law with a small power \cite{Martin:1980rm}, and as
predicted in the quark model of Ref.\ \cite{Godfrey:1985xj}.  (Ref.\
\cite{Zhang:2006yj} prefers to identify this state as the lowest $^3D_1$ $c
\bar s$ level on the basis of the $^3P_0$ decay model mentioned earlier.)

An even higher-lying $c \bar s$ state has been observed by the BaBar
Collaboration \cite{Aubert:2006mh}.  It is seen decaying to $D^0 K^+$ and
$D^+ K_S$, so it must have natural spin-parity $0^+$, $1^-$, $2^+, \ldots$.
Its mass and width are $(2856.6 \pm 1.5 \pm 5.0)$ and $(48 \pm 7 \pm 10)$
MeV/$c^2$.  It has been interpreted as a radial excitation of the $0^+$
state $D_{s0}(2317)$, i.e. the $0^+(2^3P_0)$ state \cite{vanBeveren:2006st,%
Close:2006gr}, shown hypothetically in Fig.\ \ref{fig:ds}, or a $3^-(^3D_3)$
state \cite{Colangelo:2006rq}.  Both
assignments are consistent with the $^3P_0$ decay model \cite{Zhang:2006yj}.

\subsection{Excited charmed-nonstrange states}
\vskip -0.1in

In contrast to the lightest $0^+$, $1^+$ charmed-strange states, which are
too light to decay to $D K$ or $D^* K$, the lightest $0^+$, $1^+$
charmed-nonstrange candidates appear to be heavy enough to decay to $D \pi$
or $D^* \pi$, and thus are expected to be broad.  Heavy quark symmetry
predicts the existence of a $0^+$, $1^+$ pair with light-quark total angular
momentum and parity $j^P = 1/2^+$ decaying to $D \pi$ or $D^* \pi$,
respectively, via an S-wave.  A $1^+$, $2^+$ pair with $j^P = 3/2^+$, decaying
primarily via a D-wave to $D^* \pi$ or both $D \pi$ and $D^* \pi$,
respectively, has been well-represented for a number of years by states at
$2422.3 \pm 1.3$ MeV/$c^2$ and $2461.1 \pm 1.6$ MeV/$c^2$ \cite{PDG}.
The situation with regard to the $j^P = 1/2^+$ candidates is much less
well-understood.  CLEO \cite{Anderson:1999wn} and Belle \cite{Abe:2003zm}
find a broad $1^+$ candidate in the range 2420--2460 MeV/$c^2$, while
Belle and FOCUS \cite{Link:2003bd} find broad $0^+$ candidates near 2300
and 2400 MeV/$c^2$, respectively.  For further details, see the review of
Ref.\ \cite{Colangelo:2006aa}.

\subsection{$D^+$ and $D_s$ decay constants}
\vskip -0.1in

CLEO has reported the first significant measurement of the $D^+$ decay
constant: $f_{D^+} = (222.6 \pm 16.7^{+2.8}_{-3.4})$ MeV \cite{Artuso:2005ym}.
This is consistent with lattice predictions, including one \cite{Aubin:2005ar}
of $(201 \pm 3 \pm 17)$ MeV.  The accuracy of the previous world average
\cite{PDG} $f_{D_s} = (267 \pm 33)$ MeV has been improved by a BaBar value
$f_{D_s} = 283 \pm 17 \pm 7 \pm 14$ MeV \cite{Aubert:2006sd} and a new
CLEO value $f_{D_s} = 280.1 \pm 11.6 \pm 6.0$ MeV
\cite{Stone06}.  The latter,
when combined with CLEO's $f_D$, leads to $f_{D_s}/f_D = 1.26 \pm 0.11 \pm
0.03$.  A lattice prediction for $f_{D_s}$ \cite{Aubin:2005ar} is $f_{D_s} =
249 \pm 3 \pm 16$ MeV, leading to $f_{D_s}/f_D = 1.24 \pm 0.01 \pm 0.07$.
One expects $f_{B_s}/f_B \simeq f_{D_s}/f_D$ so better measurements
of $f_{D_s}$ and $f_D$ by CLEO will help validate lattice calculations and
provide input for interpreting $B_s$ mixing.  A desirable error on $f_{B_s}/f_B
\simeq f_{D_s}/f_D$ is $\le 5\%$ for useful determination of CKM element ratio
$|V_{td}/V_{ts}|$, needing errors $\le 10$ MeV on $f_{D_s}$ and $f_D$.
The ratio $|V_{td}/V_{ts}| = 0.2060 \pm 0.0007~({\rm exp})^{+0.0081}_{-0.0080}
~({\rm theor})$ is implied by a recent CDF result on $B_s$--$\overline{B}_s$
mixing \cite{Abulencia:2006ze} combined with
$B$--$\overline{B}$ mixing and $\xi \equiv (f_{B_s} \sqrt{B_{B_s}}/f_B
\sqrt{B_B}) = 1.21^{+0.047}_{-0.035}$ from the lattice \cite{Okamoto}.
A simple quark model scaling argument anticipated $f_{D_s}/f_D \simeq
f_{B_s}/f_B \simeq \sqrt{m_s/m_d} \simeq 1.25$, where $m_s \simeq 485$ MeV and
$m_d \simeq 310$ MeV are constituent quark masses \cite{Rosner90}.

\section{BEAUTY/BOTTOM HADRONS \label{sec:beauty}}

It has become common to refer to the quantum number $B$ possessed by the
bottom quark $b$ as ``beauty.''  In accord with the convention whereby the
strangeness $S$ of the $s$ quark is assigned to be $S(s) = -1$, we take
$B(b) = -1$, $B(\bar b) = +1$.

\begin{figure}
\includegraphics[width=0.98\textwidth]{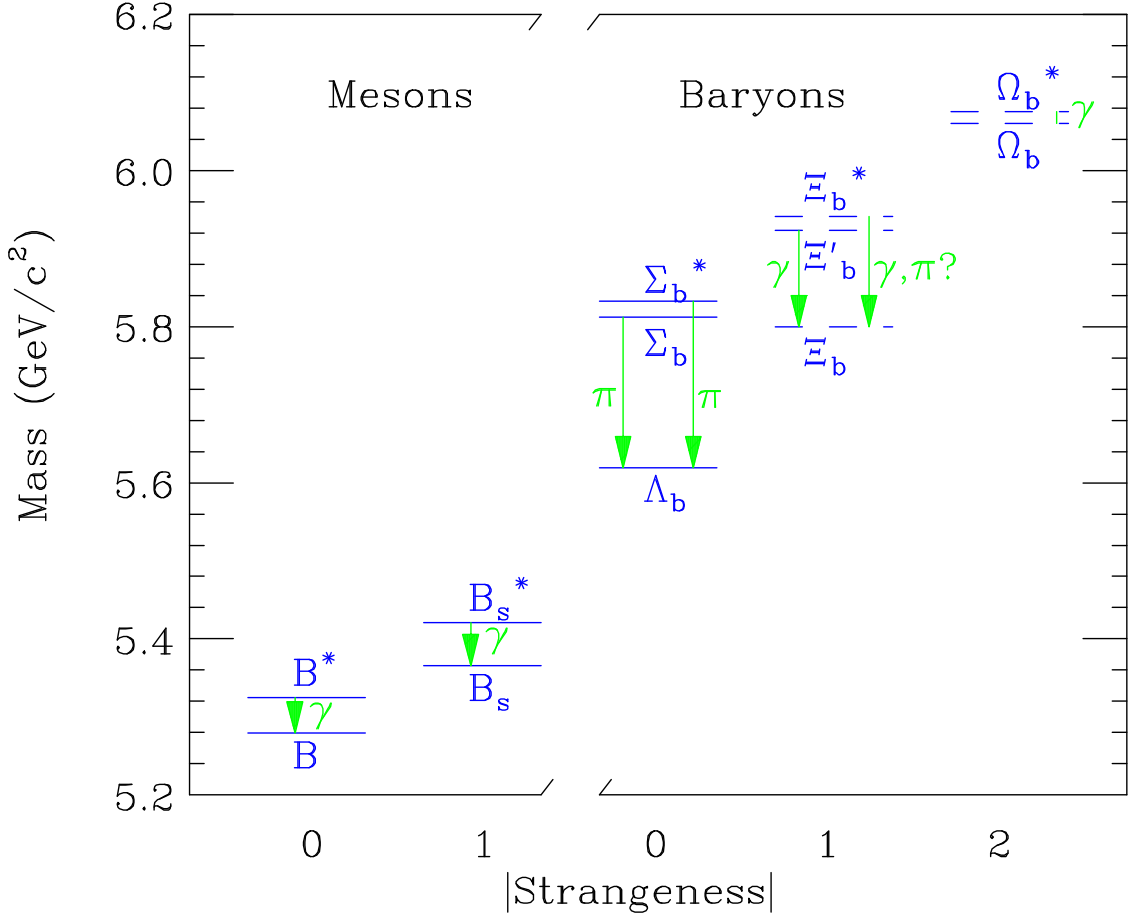}
\caption{S-wave hadrons containing a single beauty quark.  Dashed lines denote
predicted levels not yet observed.
\label{fig:beauty}}
\end{figure}

The spectrum of ground-state hadrons containing a single $b$ quark is shown in
Fig.\ \ref{fig:beauty}.  The CDF Collaboration has published measurements of
the $B_s$ and $\Lambda_b$ masses and the $B_s$--$B^0$ and $\Lambda_b$--$B^0$
mass differences which are of better precision than the current world averages
\cite{Acosta:2005mq}, and now have nearly five times the data on which that
publication was based.  With the current data sample of 1 fb$^{-1}$ CDF now
has evidence for the long-sought $\Sigma_b$ and $\Sigma^*_b$ states
\cite{Gorelov:2007ry} very near the masses predicted from the corresponding
charmed baryons using heavy quark symmetry \cite{Jenkins:1996de} or simple
quark-model ideas \cite{Karliner:2003sy}.  Using measured mass differences and
their value \cite{Acosta:2005mq} of $M(\Lambda_b) = 5619.7 \pm 1.2 \pm 1.2$
MeV/$c^2$, CDF reports
\bea \nonumber
M(\Sigma_b^-) & = & 5815.2^{~+~1.0}_{~-~0.9} \pm 1.7~{\rm MeV}/c^2~~,~~~
M(\Sigma_b^+) = 5807.5^{~+~1.9}_{~-~2.2} \pm 1.7~{\rm MeV}/c^2~~,~~~ \\
M(\Sigma_b^{*-}) & = & 5836.7^{~+~2.0}_{~-~2.3} \pm 1.7~{\rm MeV}/c^2~~,~~~
M(\Sigma_b^{*+}) = 5829.0^{~+~1.6}_{~-~1.7} \pm 1.7~{\rm MeV}/c^2~~.
\eea
The analysis employs the assumption $M(\Sigma_b^{*-}) - M(\Sigma_b^{*+}) =
M(\Sigma_b^-) - M(\Sigma_b^+)$, which was found in Ref.\ \cite{Rosner:2006yk}
to be valid to 0.4 MeV/$c^2$.  The systematic error is dominated by uncertainty
in $M(\Lambda_b)$.

The CDF Collaboration has identified events of the form $B_c \to J/\psi
\pi^\pm$, allowing a precise determination of the
mass: $M$=(6276.5$\pm$4.0$\pm$2.7) MeV/$c^2$ \cite{Aoki:2006}.  This is in
reasonable accord with the latest lattice prediction of
6304$\pm$12$^{+18}_{-0}$ MeV \cite{Allison:2004be}.

The long-awaited $B_s$--$\overline{B}_s$ mixing has finally been observed
\cite{Abulencia:2006ze,D0mix}. The CDF value, $\Delta m_s = 17.77\pm0.10\pm
0.07$ ps$^{-1}$, constrains $f_{B_s}$ and $|V_{td}/V_{ts}|$, as mentioned
earlier.

The Belle Collaboration has observed the decay $B \to \tau \nu_\tau$
\cite{Btaunu}, leading to $f_B |V_{ub}| = (10.1^{+1.6+1.1}_{-1.4-1.3})
\times 10^{-4}$ GeV.  When combined with the value $|V_{ub}| = (4.39 \pm 0.33)
\times 10^{-3}$ \cite{HFAG}, this leads to $f_B = (229^{+36+30}_{-31-34})$
MeV.  A recent unquenched lattice estimate \cite{Gray:2005ad} is $f_B =
(216\pm22)$ MeV.

A new CDF value for the $\Lambda_b$ lifetime, $\tau(\Lambda_b) = (1.593
^{+0.083}_{-0.078} \pm 0.033)$ ps, was reported recently
\cite{Abulencia:2006dr}.  Whereas
the previous world average of $\tau(\Lambda_b)$ was about 0.8 that of $B^0$,
below theoretical predictions, the new CDF value substantially increases the
world average to a value $\tau(\Lambda_b) = (1.410 \pm 0.054)$ ps which is
$0.923 \pm 0.036$ that of $B^0$ and quite comfortable with theory.

\section{CHARMONIUM \label{sec:charmon}}

\subsection{Observation of the $h_c$}
\vskip -0.1in

The $h_c(1^1P_1)$ state of charmonium has been observed by CLEO
\cite{Rosner:2005ry,Rubin:2005px} via $\psi(2S) \to \pi^0 h_c$ with $h_c
\to \gamma \eta_c$.
Hyperfine splittings test the spin-dependence and spatial behavior of the $Q
\bar Q$ force.  Whereas these splittings are $M(J/\psi) - M(\eta_c) \simeq 115$
MeV for 1S and $M[\psi'] - M(\eta'_c) \simeq $49 MeV for 2S levels, P-wave
splittings should be less than a few MeV since the potential is proportional to
$\delta^3(\vec{r})$ for a Coulomb-like $c \bar c$ interaction.  Lattice QCD
\cite{latt} and relativistic potential \cite{Ebert:2002pp} calculations confirm
this expectation.  One expects $M(h_c) \equiv M(1^1P_1) \simeq
\langle M(^3P_J) \rangle = 3525.36 \pm 0.06$ MeV.

Earlier $h_c$ sightings \cite{Rosner:2005ry,Rubin:2005px} based on
$\bar p p$ production in the direct channel, include a few events at $3525.4
\pm 0.8$ MeV seen in CERN ISR Experiment R704; a state at $3526.2 \pm
0.15 \pm 0.2$ MeV, decaying to $\pi^0 J/\psi$, reported by Fermilab E760 but
not confirmed by Fermilab E835; and a state at $3525.8 \pm 0.2 \pm 0.2$ MeV,
decaying to $\gamma \eta_c$ with $\eta_c \to \gamma \gamma$, reported by
E835 with about a dozen candidate events \cite{Andreotti:2005vu}.

\begin{figure}
\mbox{
\includegraphics[width=0.58\textwidth]{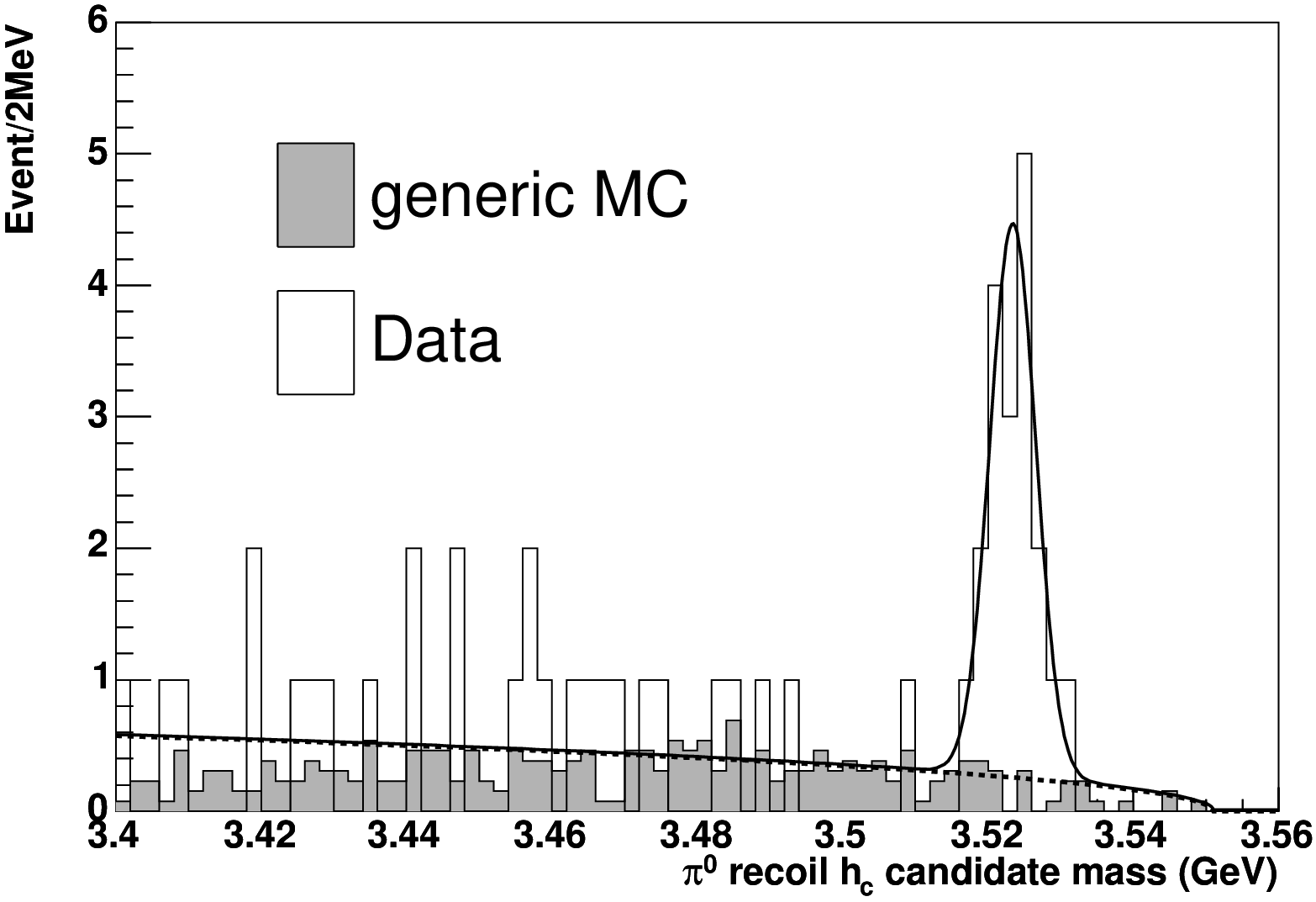}
\includegraphics[width=0.41\textwidth]{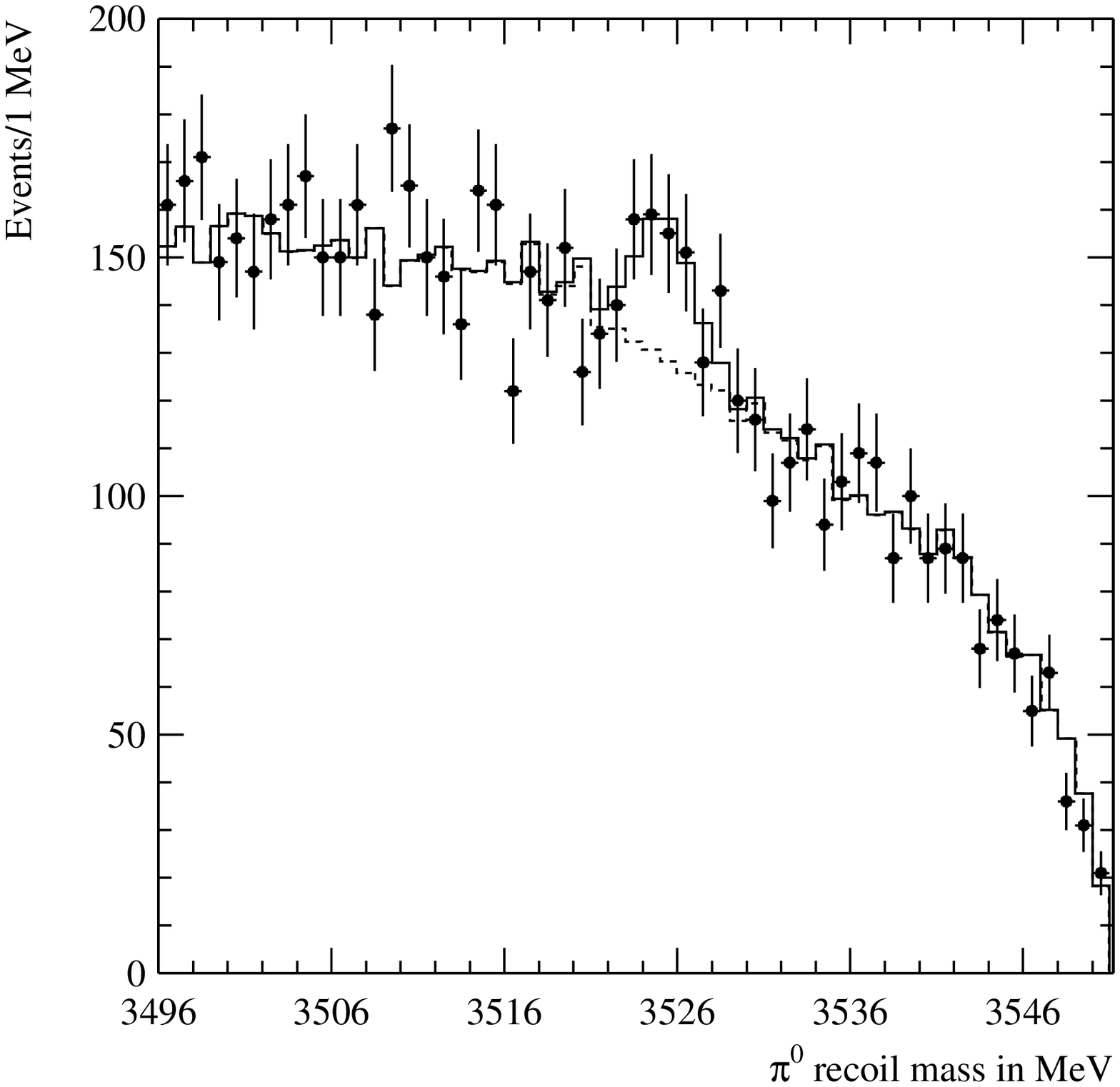}
}
\caption{Left: Exclusive $h_c$ signal from CLEO (3 million $\psi(2S)$
decays).  Data events correspond to open histogram;
Monte Carlo background estimate is denoted by shaded histogram.
The signal shape is a double Gaussian, obtained from signal Monte Carlo.
The background shape is an ARGUS function.
Right: Inclusive $h_c$ signal from CLEO (3 million $\psi(2S)$ decays).
The curve denotes the background function based on generic Monte Carlo plus
signal.  The dashed line shows the contribution of background alone.
Both figures are from Ref.\ \cite{Rubin:2005px}.
\label{fig:hc}}
\end{figure}

In the CLEO data, both inclusive and exclusive analyses see a signal near
$\langle M(^3P_J) \rangle$.  The exclusive analysis reconstructs $\eta_c$ in 7
decay modes, while no $\eta_c$ reconstruction is performed in the inclusive
analysis.
The exclusive signal is shown on the left in Fig.\ \ref{fig:hc}.  A total of 19
candidates were identified, with a signal of $17.5 \pm 4.5$ events above
background.  The mass and product branching ratio for the two transitions 
are $M(h_c) = (3523.6 \pm 0.9 \pm 0.5)$ MeV; ${\cal B}_1(\psi' \to \pi^0 h_c)
{\cal B}_2(h_c \to \gamma \eta_c) = (5.3 \pm 1.5 \pm 1.0) \times 10^{-4}$.
The result of one of two inclusive analyses is shown on the right in
Fig.\ \ref{fig:hc}.  These yield $M(h_c) = (3524.9 \pm 0.7 \pm 0.4)$ MeV,
${\cal B}_1 {\cal B}_2 = (3.5 \pm 1.0 \pm 0.7) \times 10^{-4}$.  Combining
exclusive and inclusive results yields $M(h_c) = (3524.4 \pm 0.6 \pm 0.4)$ MeV,
${\cal B}_1 {\cal B}_2 = (4.0 \pm 0.8 \pm 0.7) \times 10^{-4}$.  The $h_c$ mass
is $(1.0 \pm 0.6 \pm 0.4)$ MeV below $\langle M(^3P_J) \rangle$, barely
consistent with the (nonrelativistic) bound \cite{Stubbe:1991qw} $M(h_c) \ge
\langle M(^3P_J) \rangle$ and indicating little P-wave hyperfine splitting in
charmonium.  The value of ${\cal B}_1 {\cal B}_2$ agrees with theoretical
estimates of $(10^{-3} \cdot 0.4)$.
\vskip -0.3in

\subsection{Decays of the $\psi'' \equiv \psi(3770)$}

The $\psi''(3770)$ is a potential ``charm factory'' for present and future $e^+
e^-$ experiments.  At one time $\sigma(e^+ e^- \to \psi'')$ seemed
larger than $\sigma(e^+ e^- \to \psi'' \to D \bar D)$, raising the question
of whether there were significant non-$D \bar D$ decays of the $\psi''$
\cite{Rosner:2004wy}.  A new CLEO measurement \cite{CLDDbar}, $\sigma(\psi'') =
(6.38 \pm 0.08 ^{+0.41}_{-0.30})$ nb, appears very close to the CLEO value
$\sigma(D \bar D) = (6.39\pm0.10^{+0.17}_{-0.08})$ nb \cite{He:2005bs}, leaving
little room for non-$D \bar D$ decays.  (In recent BES analyses \cite{BESsig}
a significant non-$D \bar D$ component could still be present.)

One finds that ${\cal B}(\psi''\to \pi \pi J/\psi,~\gamma \chi_{cJ}, \ldots)$
sum to at most 1--2\%.  Moreover, both CLEO and BES \cite{LP123}, in searching
for enhanced light-hadron modes, find only that the $\rho \pi$ mode,
suppressed in $\psi(2S)$ decays, also is {\it suppressed} in $\psi''$ decays.

Some branching ratios for $\psi'' \to X J/\psi$ \cite{Adam:2005mr} are
${\cal B}(\psi'' \to \pi^+ \pi^- J/\psi) =(0.189\pm0.020\pm0.020)\%$,
${\cal B}(\psi'' \to \pi^0 \pi^0 J/\psi) =(0.080\pm0.025\pm0.016)\%$,
${\cal B}(\psi'' \to \eta J/\psi) = (0.087\pm0.033\pm0.022)\%$, and
${\cal B}(\psi'' \to \pi^0 J/\psi) < 0.028\%$.
The value of ${\cal B}[\psi''(3770) \to \pi^+ \pi^- J/\psi]$ found by CLEO is a
bit above 1/2 that reported by BES \cite{Bai:2003hv}.
These account for less than 1/2\% of the total $\psi''$ decays.

\begin{table}
\caption{CLEO results on radiative decays $\psi'' \to \gamma \chi_{cJ}$.
Theoretical predictions of \cite{Eichten:2004uh} are (a) without and
(b) with coupled-channel effects; (c) shows predictions of
\cite{Rosner:2004wy}.
\label{tab:psipprad}}
\begin{center}
\begin{tabular}{ccccc} \hline
Mode & \multicolumn{3}{c}{Predicted (keV)} & CLEO \\
     & (a) & (b) & (c) & \cite{Briere:2006ff} \\ \hline
$\gamma \chi_{c2}$ & 3.2 & 3.9 & 24$\pm$4 & $<21$ \\
$\gamma \chi_{c1}$ & 183 & 59 & $73\pm9$ & $75\pm18$ \\
$\gamma \chi_{c0}$ & 254 & 225 & 523$\pm$12 & $172\pm30$ \\ \hline
\end{tabular}
\end{center}
\end{table}

CLEO has reported results on $\psi'' \to \gamma \chi_{cJ}$ partial
widths, based on the exclusive process $\psi'' \to \gamma \chi_{c1,2} \to
\gamma \gamma J/\psi \to \gamma \gamma \ell^+ \ell^-$ \cite{Coan:2005} and
reconstruction of exclusive $\chi_{cJ}$ decays \cite{Briere:2006ff}.  The
results are shown in Table \ref{tab:psipprad}, implying
$\sum_J{\cal B}(\psi'' \to \gamma \chi_{cJ}) = {\cal O}$(1\%).

Several searches for $\psi''(3770) \to ({\rm light~ hadrons})$, including VP,
$K_L K_S$, and multi-body final states have been performed.  Two CLEO analyses
\cite{Adams:2005ks,Huang:2005} find no evidence for any light-hadron $\psi''$
mode above expectations from continuum production except $\phi \eta$,
indicating no obvious signature of non-$D \bar D$ $\psi''$ decays.
\vskip -0.3in

\subsection{$X(3872)$: A $1^{++}$ molecule}
\vskip -0.1in

Many charmonium states above $D \bar D$ threshold have been seen recently.
Reviews may be found in Refs.\ \cite{GodfreyFPCP,Swanson}.  The $X(3872)$,
discovered by Belle in $B$ decays \cite{Choi:2003ue} and confirmed by BaBar
\cite{Aubert:2004ns} and in hadronic production \cite{Acosta:2003zx}, was seen
first in its decay to $J/\psi \pi^+ \pi^-$.  Since it lies well
above $D \bar D$ threshold but is narrower than experimental resolution (a few
MeV), unnatural $J^P = 0^-,1^+, 2^-$ is favored.  It has many features in
common with an S-wave bound state of $(D^0 \bar D^{*0} + \bar D^0 D^{*0})/
\sqrt{2} \sim c \bar c u \bar u$ with $J^{PC} = 1^{++}$ \cite{Close:2003sg}.
The simultaneous decay of $X(3872)$ to $\rho J/\psi$ and $\omega J/\psi$ with
roughly equal branching ratios is a consequence of this ``molecular''
assignment.

Analysis of angular distributions \cite{Rosner:2004ac} in $X \to \rho J/\psi,
\omega J/\psi$ favors the $1^{++}$ assignment \cite{Abe:2005iy}.  (See also
\cite{Marsiske,Swanson}.) Although a high-statistics analysis by CDF cannot
distinguish between $J^{PC} = 1^{++}$ or $2^{-+}$ \cite{Abulencia:2006ma}, the
latter assignment is disfavored by Belle's observation \cite{Gokhroo:2006bt} of
$X \to D^0 \bar D^0 \pi^0$, which would require at least two units of relative
orbital angular momentum in the three-body state, very near threshold.

The detection of the $\gamma J/\psi$ mode ($\sim
14\%$ of $J/\psi \pi^+ \pi^-$) \cite{Abe:2005ix} confirms the assignment of
positive $C$ and suggests a $c \bar c$ admixture in the wave function.  BaBar
\cite{Bapipipsi} finds ${\cal B}[X(3872) \to \pi^+ \pi^- J/\psi] > 0.042$ at
90\% c.l.

\subsection{Additional states around 3940 MeV}
\vskip -0.1in

Belle has reported a candidate for a $2^3P_2(\chi'_{c2})$ state in $\gamma
\gamma$ collisions \cite{Abe:2005bp}, decaying to $D \bar D$.
The angular distribution of $D \bar D$ pairs is
consistent with $\sin^4 \theta^*$ as expected for a state with $J=2, \lambda =
\pm2$.  It has $M = 3929 \pm 5 \pm 2$ MeV, $\Gamma = 29 \pm 10 \pm 3$ MeV, and
$\Gamma_{ee} {\cal B}(D \bar D) = 0.18 \pm 0.06 \pm 0.03$ eV, all reasonable
for a $\chi'_{c2}$ state.

A charmonium state $X(3938)$ is produced recoiling against $J/\psi$ in $e^+ e^-
\to J/\psi + X$ \cite{Abe:2005hd} and is seen decaying to $D \bar D^*$ +
c.c.  Since all lower-mass states observed in this recoil process have $J=0$
(these are the $\eta_c(1S), \chi_{c0}$ and $\eta'_c(2S)$), it is tempting to
identify this state with $\eta_c(3S)$ (not $\chi'_{c0}$, which would decay to
$D \bar D$).

The $\omega J/\psi$ final state in $B \to K \omega J/\psi$ shows a peak above
threshold at $M(\omega J/\psi) \simeq 3940$ MeV \cite{Abe:2004zs}.  This could
be a candidate for one or more excited P-wave charmonium states, likely the
$\chi'_{c1,2}(2^3P_{1,2})$.  The corresponding $b \bar b$ states $\chi'_{b1,2}$
have been seen to decay to $\omega \Upsilon(1S)$ \cite{Severini:2003qw}.
\vskip -0.3in

\subsection{The $Y(4260)$}
\vskip -0.1in

\begin{figure}
\includegraphics[height=0.46\textheight]{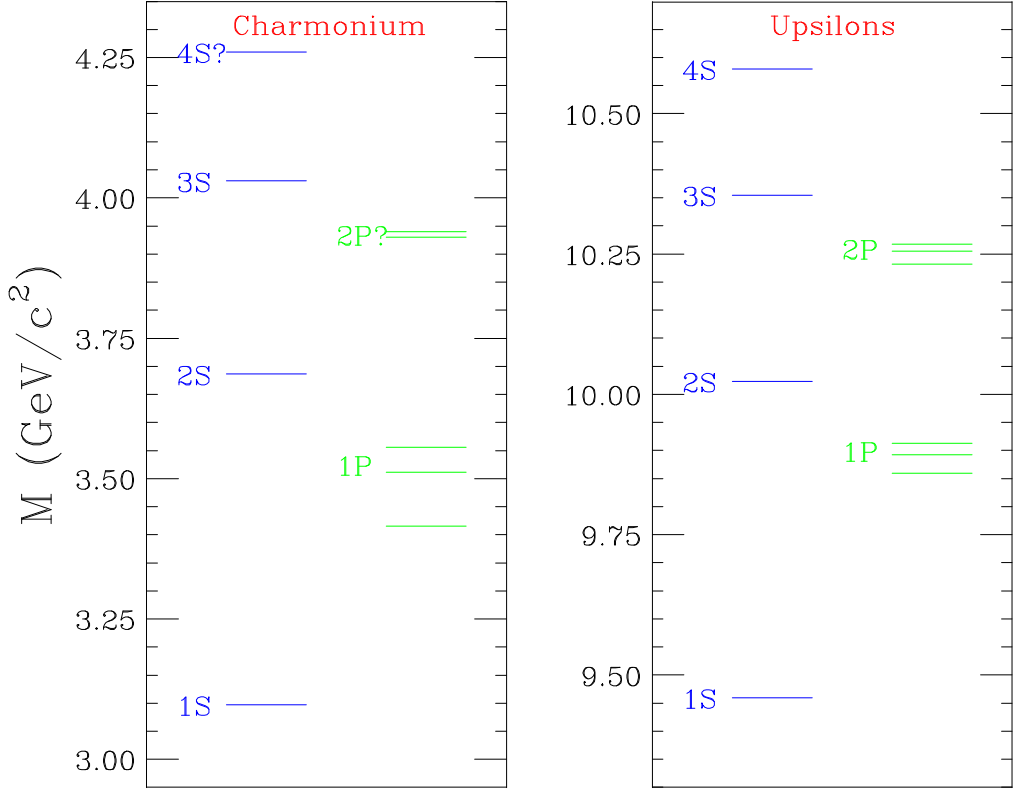}
\caption{Congruence of charmonium and bottomonium spectra if the $Y(4260)$
is a 4S level.
\label{fig:comp}}
\end{figure}

Last year BaBar reported a state $Y(4260)$ produced in the radiative return
reaction $e^+ e^- \to \gamma \pi^+ \pi^- J/\psi$ and seen in the $\pi^+ \pi^-
J/\psi$ spectrum \cite{Aubert:2005rm}.  Its mass is consistent with being a
$4S$ level \cite{Llanes-Estrada:2005vf} since it lies about 230 MeV above the
$3S$ candidate (to be compared with a similar $4S$-$3S$ spacing in the
$\Upsilon$ system).  Indeed, a $4S$ charmonium level at 4260 MeV/$c^2$ was
anticipated on this basis \cite{Quigg:1977dd}.  With this assignment,
the $nS$ levels of charmonium and bottomonium are nearly congruent to one
another, as shown in Fig.\ \ref{fig:comp}.  Their spacings would be identical
if the interquark potential were $V(r) \sim {\rm log}(r)$, which may be viewed
as an interpolation between the short-distance $\sim -1/r$ and long-distance
$\sim r$ behavior expected in QCD.  Circumstances weighing
against the $4S$ assignment include the lack of a peak in $R \equiv
\sigma(e^+ e^- \to {\rm hadrons})/\sigma(e^+ e^- \to \mu^+ \mu^-)$ at 4260
MeV ($R$ exhibits a {\it dip} just below this energy), and a $4S$
candidate at 4415 MeV/$c^2$ in accord with many potential models.

The CLEO Collaboration has confirmed the $Y(4260)$, both in a direct scan
\cite{Coan:2006rv} and in radiative return \cite{Blusk}.  Signals are seen
for $Y(4260)
\to \pi^+ \pi^- J/\psi$ (11$\sigma$), $\pi^0 \pi^0 J/\psi$ (5.1$\sigma$), and
$K^+ K^- J/\psi$ (3.7$\sigma$).  There are also weak signals for $\psi(4160)
\to \pi^+ \pi^- J/\psi$ (3.6$\sigma$) and $\pi^0 \pi^0 J/\psi$ (2.6$\sigma$),
consistent with the $Y(4260)$ tail, and for $\psi(4040) \to \pi^+ \pi^- J/\psi$
(3.3$\sigma$).

Other interpretations of $Y(4260)$ include a $c s \bar c \bar s$ state
\cite{Maiani:2005pe} and a hybrid $c \bar c g$ state \cite{Zhu:2005hp}, for
which it lies in the expected mass range.  One consequence of the hybrid
interpretation is a predicted decay to $D \bar D_1 +$ c.c., where $D_1$
is a P-wave $c \bar q$ pair.  Now, $D \bar D_1$ threshold is 4287 MeV/$c^2$
if we consider the lightest $D_1$ to be the state noted in Ref.\ \cite{PDG}
at 2422 MeV/$c^2$.  In this case the $Y(4260)$ would be a $D \bar D_1 +$ c.c.
{\it bound state}.  It would decay to $D \pi \bar D^*$, where the $D$ and $\pi$
are not in a $D^*$. The dip in $R_{e^+ e^-}$ lies just below $D \pi \bar
D^*$ threshold, which may be the first S-wave meson pair accessible in
$c \bar c$ fragmentation \cite{Close:2005iz}.

\subsection{Charmonium: updated}

\begin{figure}
\includegraphics[width=0.98\textwidth]{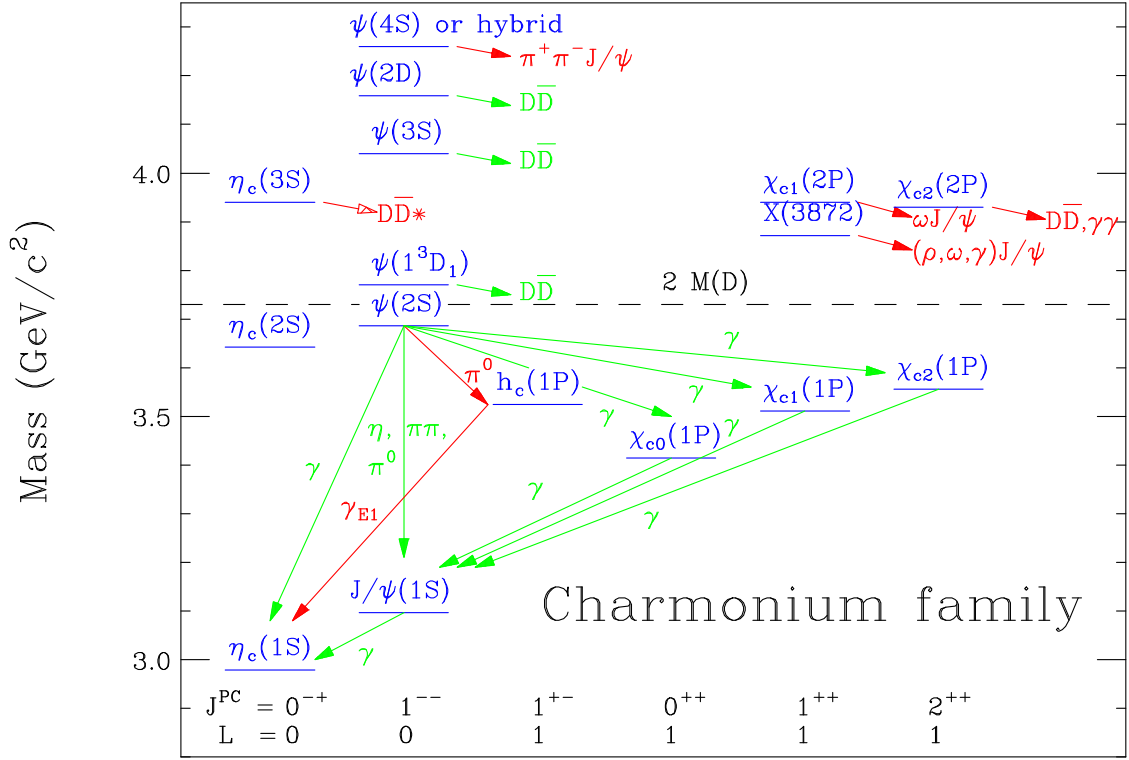}
\caption{Charmonium states including levels above charm threshold.
\label{fig:charmon}}
\end{figure}

Remarkable progress has been made in the spectroscopy of charmonium states
above charm threshold in the past few years.  Fig.\ \ref{fig:charmon}
summarizes the levels (some of whose assignments are tentative).  Even though
such states can decay to charmed pairs (with the possible exception of
$X(3872)$, which may be just below $D \bar D^*$ threshold), other decay modes
are being seen.

\section{THE $\Upsilon$ FAMILY (BOTTOMONIUM) \label{sec:upsilons}}

\begin{figure}
\includegraphics[height=0.48\textheight]{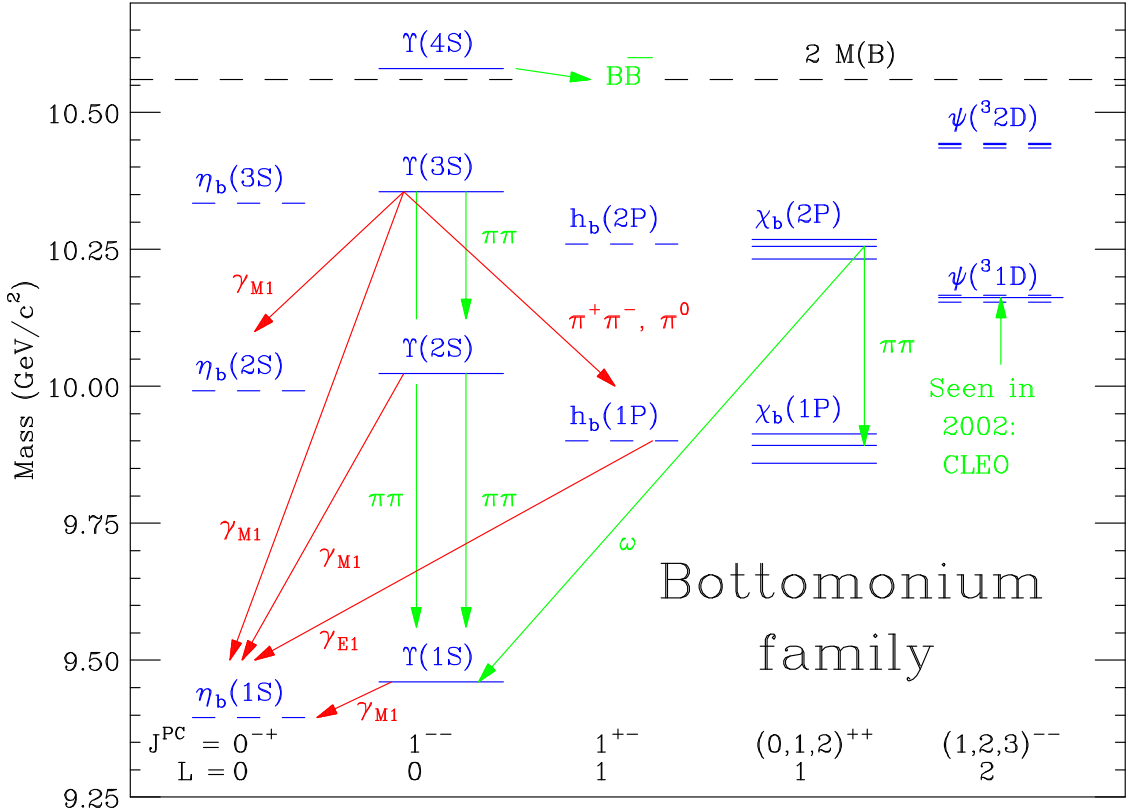}
\caption{Bottomonium levels and some decays.  Electric dipole (E1) transitions
$S \leftrightarrow P \leftrightarrow D$ are not shown.
\label{fig:ups}}
\end{figure}

Some properties and decays of the $\Upsilon$ ($b \bar b$) levels are summarized
in Fig. \ref{fig:ups}.  Masses are in agreement with unquenched lattice QCD
calculations \cite{Lepage}.  Direct photons have been
observed in 1S, 2S, and 3S decays, implying estimates of the strong
fine-structure constant consistent with others \cite{Besson:2005jv}.  The
transitions $\chi_b(2P) \to \pi \pi \chi_b(1P)$ have been seen
\cite{Cawlfield:2005ra,Tati}.  BaBar has measured the partial widths
$\Gamma[\Upsilon(4S) \to \pi^+ \pi^- \Upsilon(1S)] = 1.8 \pm 0.4$ keV and
$\Gamma[\Upsilon(4S) \to \pi^+ \pi^- \Upsilon(2S)] = 2.7 \pm 0.8$ keV
\cite{Aubert:2006bm}, while Belle
has seen $\Upsilon(4S) \to \pi^+ \pi^- \Upsilon(1S)$, with a branching ratio
${\cal B} = (1.1 \pm 0.2 \pm 0.4) \times 10^{-4}$ \cite{BeUps}.

\subsection{Remeasurement of $\Upsilon(nS)$ properties}
\vskip -0.1in

New values of ${\cal B}[\Upsilon(1S,2S,3S) \to \mu^+ \mu^-] = (2.49 \pm 0.02
\pm 0.07,~2.03\pm0.03\pm0.08,~2.39\pm0.07\pm0.10)\%$ \cite{Adams:2004xa},
when combined with new measurements $\Gamma_{ee}(1S,2S,3S) = (1.252\pm0.004
\pm0.019,~0.581\pm0.004,\pm0.009,~0.413\pm0.004\pm0.006)$ keV imply total
widths $\Gamma_{\rm tot}(1S,2S,3S) = (50.3\pm1.7,~28.6\pm1.3,~17.3\pm0.6)$ keV.
The values of $\Gamma_{\rm tot}(2S,3S)$ changed considerably with respect
to previous world averages.  Combining with previous data, the Particle Data
Group \cite{PDG} now quotes $\Gamma_{\rm tot}(1S,2S,3S)=(54.02\pm1.25,%
~31.98\pm2.63,~20.32\pm1.85)$ keV, which we shall use in what follows.  This
will lead to changes in comparisons of predicted and observed transition rates.
As one example, the study of $\Upsilon(2S,3S) \to \gamma X$ decays
\cite{Artuso:2004fp} has provided new branching ratios for E1 transitions
to $\chi_{bJ}(1P),~\chi'_{bJ}(2P)$ states.  These may be combined with the
new total widths to obtain updated partial decay widths [line (a) in
Table \ref{tab:E1}], which may be compared with one set of non-relativistic
predictions \cite{KR} [line (b)].  The suppression of transitions to $J=0$
states by 10--20\% with respect to non-relativistic expectations agrees
with relativistic predictions \cite{rel}.  The partial width for $\Upsilon(3S)
\to \gamma 1^3P_0$ is found to be $61 \pm 23$ eV, about nine times the
highly-suppressed value predicted in Ref.\ \cite{KR}.  That prediction is
very sensitive to details of wave functions; the discrepancy indicates
the importance of relativistic distortions.

\begin{table}
\caption{Comparison of observed (a) and predicted (b) partial widths
for $2S \to 1 P_J$ and $3S \to 2 P_J$ transitions in $b \bar b$ systems.
\label{tab:E1}}
\begin{center}
\begin{tabular}{|c|c c c|c c c|} \hline
 & \multicolumn{3}{c|}{$\Gamma$ (keV), $2S \to 1P_J$ transitions}
 & \multicolumn{3}{c|}{$\Gamma$ (keV), $3S \to 2P_J$ transitions} \\
 & $J=0$ & $J=1$ & $J=2$ & $J=0$ & $J=1$ & $J=2$ \\ \hline
(a) & 1.20$\pm$0.18 & 2.22$\pm$0.23 & 2.32$\pm$0.23 &
 1.38$\pm$0.19 & 2.95$\pm$0.30 & 3.21$\pm$0.33 \\
(b)& 1.39 & 2.18 & 2.14 & 1.65 & 2.52 & 2.78 \\ \hline \hline
\end{tabular}
\end{center}
\end{table}

\subsection{$b \bar b$ spin singlets}
\vskip -0.1in

Decays of the $\Upsilon(1S,2S,3S)$ states are potential sources of information
on $b \bar b$ spin-singlets, but none has been seen yet.  One expects
1S, 2S, and 3S hyperfine splittings to be approximately 60, 30, 20 MeV/$c^2$,
respectively \cite{Godfrey:2001eb}.  The lowest P-wave singlet state (``$h_b$'')
is expected to be near $\langle M(1^3P_J) \rangle \simeq 9900$ MeV/$c^2$
\cite{Godfrey:2002rp}.

Several searches have been performed or are under way in 1S, 2S, and 3S CLEO
data.  One can search for the allowed M1 transition in $\Upsilon(1S) \to \gamma
\eta_b(1S)$ by reconstructing exclusive final states in $\eta_b(1S)$ decays
and dispensing with the soft photon, which is likely to be swallowed up in
background.  Final states are likely to be of high multiplicity.

One can search for higher-energy but suppressed M1 photons in $\Upsilon(n'S)
\to \gamma \eta_b(nS)\\(n \ne n')$ decays.  These searches already exclude many
models. The strongest upper limit obtained is for $n'=3$, $n=1$: ${\cal B} \le
4.3 \times 10^{-4}$ (90\% c.l.).  $\eta_b$ searches using sequential processes
$\Upsilon(3S) \to \pi^0 h_b(1^1P_1) \to \pi^0 \gamma \eta_b(1S)$ and
$\Upsilon(3S) \to \gamma \chi'_{b0} \to \gamma \eta \eta_b(1S)$ (the latter
suggested in Ref.\ \cite{Voloshin:2004hs}) are being conducted but there are no
results yet.  Additional searches for $h_b$ involve the transition
$\Upsilon(3S) \to \pi^+ \pi^- h_b$ [for which a typical experimental upper
bound based on earlier CLEO data \cite{Brock:1990pj} is
${\cal O}(10^{-3}$)], with a possible $h_b \to \gamma \eta_b$ transition
expected to have a 40\% branching ratio \cite{Godfrey:2002rp}.

\section{FUTURE PROSPECTS \label{sec:future}}

Two main sources of information on hadron spectroscopy in the past few years
have been BES-II and CLEO. BES-II has ceased operation to make way for BES-III.
CLEO plans to focus on center-of-mass energies of 3770 and 4170 MeV, split
roughly equally, with goals of about 750 pb$^{-1}$ at each energy.  The
determination of $f_D$, $f_{D_s}$, and form factors for semileptonic $D$ and
$D_s$ decays will provide incisive tests for lattice gauge theories and measure
CKM factors $V_{cd}$ and $V_{cs}$ with unprecedented precision.  In addition,
about 25 million $\psi(2S)$ (about 8 times the current CLEO sample) have been
accumulated in the summer of 2006.  CLEO-c running will end at the end of March
2008; BES-III and PANDA will extend charm studies further.

Belle has taken 3 fb$^{-1}$ of data at $\Upsilon(3S)$ [CLEO has (1.1,1.2,1.2)
fb$^{-1}$ at (1S,2S,3S)].  Both BaBar and Belle are well-situated to study
hadron spectroscopy.  We look forward to further contributions
from CDF and D0, and to the LHCb experiment, with its detailed studies of
charm and $b$ hadrons produced in hadronic collisions.

\section{SUMMARY \label{sec:summary}}

Hadron spectroscopy is providing both long-awaited states like $h_c$ and
surprises like low-lying P-wave $D_s$ mesons, X(3872), X(3940), Y(3940),
Z(3940) and Y(4260).  Decays of $\psi''(3770)$ shed light on its nature.
Higher excitations
of charmed baryons and charmed-strange mesons continue to be uncovered.
Upon reflection, some properties of the new hadron states may be less
surprising but we are continuing to learn about properties of QCD in the
strong-coupling regime.  There is evidence for molecules, 3S, 2P, 4S or hybrid
charmonium, and interesting decays of states above flavor threshold.

QCD may not be the last strongly coupled theory with which we have to deal.
The mystery of electroweak symmetry breaking may require related techniques,
if the long-awaited Higgs boson is not elementary but a phenomenon emerging
from a new strongly-interacting sector (as in the ``technicolor'' scheme or
other pictures of composite Higgs bosons).  Understanding the very structure of
quarks and leptons may require spectroscopic techniques.  This will be a very
different spectroscopy than that of hadrons, since all evidence points to a
large separation between the masses of the
known elementary fermions and any possible compositeness scale.  The insights
on hadron spectra described here are coming to us in general from experiments
at the frontier of intensity and detector capabilities rather than energy,
illustrating the importance of a diverse approach to the fundamental
structure of matter.

\section*{ACKNOWLEDGMENTS}

I thank colleagues on BaBar, Belle, and CLEO for sharing data and for helpful
discussions, and E. van Beveren, D. Bugg, F. Close, L. Glozman, J. Hedditch, J.
Pel\'aez, and G. Rupp for some helpful references.  This work was supported
in part by the United States Department of Energy under Grant No.\
DE FG02 90ER40560.

\end{document}